\documentclass[twocolumn,showpacs,preprintnumbers,amsmath,amssymb]{revtex4}

\usepackage[T1]{fontenc}
\usepackage{graphicx}
\usepackage{dcolumn}
\usepackage{bm}

\begin{document}

\title{A Novel Approach to Model Hybrid Stars}

\author{V.A. Dexheimer}
 \email{dexheimer@th.physik.uni-frankfurt.de}
\affiliation{FIAS, Johann Wolfgang Goethe University, Frankfurt am Main, Germany}

\author{S. Schramm}
 \email{schramm@th.physik.uni-frankfurt.de}
\affiliation{CSC, FIAS, ITP, Johann Wolfgang Goethe University, Frankfurt am Main, Germany}

\date{\today}

\begin{abstract}
We extend the hadronic SU(3) non-linear sigma model to include
quark degrees of freedom. The choice of potential for the
deconfinement order parameter as a function of temperature and
chemical potential allows us to construct a realistic phase
diagram from the analysis of the order parameters of the
system. These parameters are the chiral condensate, for the
chiral symmetry restoration, and the scalar field $\Phi$ (as an
effective field related to the Polyakov loop) for the
deconfinement to quark matter. Besides reproducing lattice QCD
results, for zero and low chemical potential, we are in
agreement with neutron star observations for zero temperature.
\end{abstract}

\maketitle

The models used to describe neutron stars can generally be divided into two classes.
The first class includes approaches in which the constituent particles are
hadrons \cite{Glendenning:1991ic,Weber:1989uq,Schaffner:1995th}. Some of them incorporate
certain symmetries from QCD, like chiral symmetry, but they do not include deconfinement.
Examples of these are  hadronic sigma
models \cite{chiral2,Heide:1993yz,Carter:1995zi,Bonanno:2008tt}. The second class includes quark star models,
which usually do not directly incorporate hadronic degrees of freedom in the model formulation.
Examples of these are bag-model studies \cite{bag0} as well as quark-NJL model  and quark sigma-models \cite{Bub}.

Using these approaches hybrid neutron stars, which consist of a hadronic and a quark phase,
are normally described by adopting
two different models with separate equations of state for hadronic and quark matter (see e.g. \cite{hybrid1}).
They are connected at the chemical potential in which the pressure of the quark EOS exceeds the hadronic one,
signalling the phase transition to quark matter. Within our approach
we employ a single model for the hadronic and for the quark phase.

The extension of the hadronic SU(3) non-linear sigma model to
quark degrees of freedom is constructed in a spirit similar to
the PNJL model \cite{PNJL}, in the sense that it is a
non-linear sigma model that introduces a scalar field which
suppress the quark degrees of freedom at low
densities/temperatures. In QCD this scalar field was named
Polyakov loop and is defined via $\Phi=\frac13$Tr$[\exp{(i\int
d\tau A_4)}]$, where $A_4=iA_0$ is the temporal component of
the SU(3) gauge field. In our case, this scalar field is also
called $\Phi$, in analogy to the PNJL approach with an
effective potential for the field, as discussed below, that
drives the phase transition in the field $\Phi$ representing  a
phenomenological description of the transition from the
confined to the deconfined phase.

The Lagrangian density of the non-linear sigma model in mean field approximation reads:
\begin{eqnarray}
&L = L_{Kin}+L_{Int}+L_{Self}+L_{SB}-U,&
\end{eqnarray}
where besides the kinetic energy term for hadrons, quarks, and leptons (included to insure charge neutrality)
the terms:
\begin{eqnarray}
&L_{Int}=-\sum_i \bar{\psi_i}[\gamma_0(g_{i\omega}\omega+g_{i\phi}\phi+g_{i\rho}\tau_3\rho)+M_i^*]\psi_i,\nonumber&\\&
\end{eqnarray}
\begin{eqnarray}
&L_{Self}=-\frac{1}{2}(m_\omega^2\omega^2+m_\rho^2\rho^2+m_\phi^2\phi^2)\nonumber&\\&
+g_4\left(\omega^4+\frac{\phi^4}{4}+3\omega^2\phi^2+\frac{4\omega^3\phi}{\sqrt{2}}+\frac{2\omega\phi^3}{\sqrt{2}}\right)\nonumber&\\&+k_0(\sigma^2+\zeta^2+\delta^2)+k_1(\sigma^2+\zeta^2+\delta^2)^2&\nonumber\\&+k_2\left(\frac{\sigma^4}{2}+\frac{\delta^4}{2}
+3\sigma^2\delta^2+\zeta^4\right)
+k_3(\sigma^2-\delta^2)\zeta&\nonumber\\&+k_4\ \ \ln{\frac{(\sigma^2-\delta^2)\zeta}{\sigma_0^2\zeta_0}},&
\end{eqnarray}
\begin{eqnarray}
&L_{SB}= m_\pi^2 f_\pi\sigma+\left(\sqrt{2}m_k^ 2f_k-\frac{1}{\sqrt{2}}m_\pi^ 2 f_\pi\right)\zeta,\nonumber&\\&
\end{eqnarray}
represent the interactions between baryons (and quarks)
and vector and scalar mesons, the self interactions of
scalar and vector mesons and an explicit chiral symmetry breaking term, responsible for producing the masses of
the pseudo-scalar mesons. The $\Phi$ potential $U$ will be discussed in the following. The underlying flavor symmetry of the
model is SU(3) and the index $i$ denotes the baryon octet and the three light quarks. The mesons included are
the vector-isoscalars $\omega$ and $\phi$, the vector-isovector $\rho$,
the scalar-isoscalars $\sigma$ and $\zeta$ (strange quark-antiquark state) and  the scalar-isovector $\delta$.
The isovector mesons affect isospin-asymmetric matter and are
consequently important for neutron star physics. The coupling constants of the model are shown in Table.~\ref{tabela1}. They were fitted to reproduce the vacuum masses of the baryons and mesons, nuclear saturation properties (density $\rho_0=0.15$ fm$^{-3}$, binding energy per nucleon $B/A=-16.00$ MeV, nucleon effective mass $M^*_N=0,67$ $M_N$, compressibility $K=297.32$ MeV), asymmetry energy ($E_{sym}=32.50$ MeV), and reasonable values for the hyperon potentials ($U_\Lambda=-28.00$ MeV, $U_\Sigma=5.35$ MeV, $U_\Xi=-18.36$ MeV). The vacuum expectation values of the scalar mesons are constrained by
reproducing the pion and kaon decay constants.
 A detailed discussion of the purely hadronic part of the Lagrangian can be found in \cite{chiral1,chiral2,eu}.

\begin{table}
\caption{\label{tabela1}Coupling constants for the model containing only baryons}
\begin{ruledtabular}
\begin{tabular}{ccc}
$ g_{N\omega}=11.90 $&$ g_{N\phi}=0 $& $ g_{N\rho}=4.03 $ \\
$ g_{N\sigma}=-9.83 $&$ g_{N\delta}=-2.34 $&$ g_{N\zeta}=1.22$ \\
$ g_{\Lambda\omega}=7.93$&$ g_{\Lambda\phi}=-7.32 $& $ g_{\Lambda\rho}=0$ \\
$ g_{\Lambda\sigma}=-5.52 $&$ g_{\Lambda\delta}=0 $&$ g_{\Lambda\zeta}=-2.30$ \\
$ k_0=1.19 $&$ k_1=-1.40 $&$ k_2=5.55 $\\
$ k_3=2.65 $&$ k_4=-0.06 $&$ g_4=38.9$ \\
\end{tabular}
\end{ruledtabular}
\end{table}

\begin{figure}[t]
\centering
\vspace{1.0cm}
\includegraphics[width=8.cm]{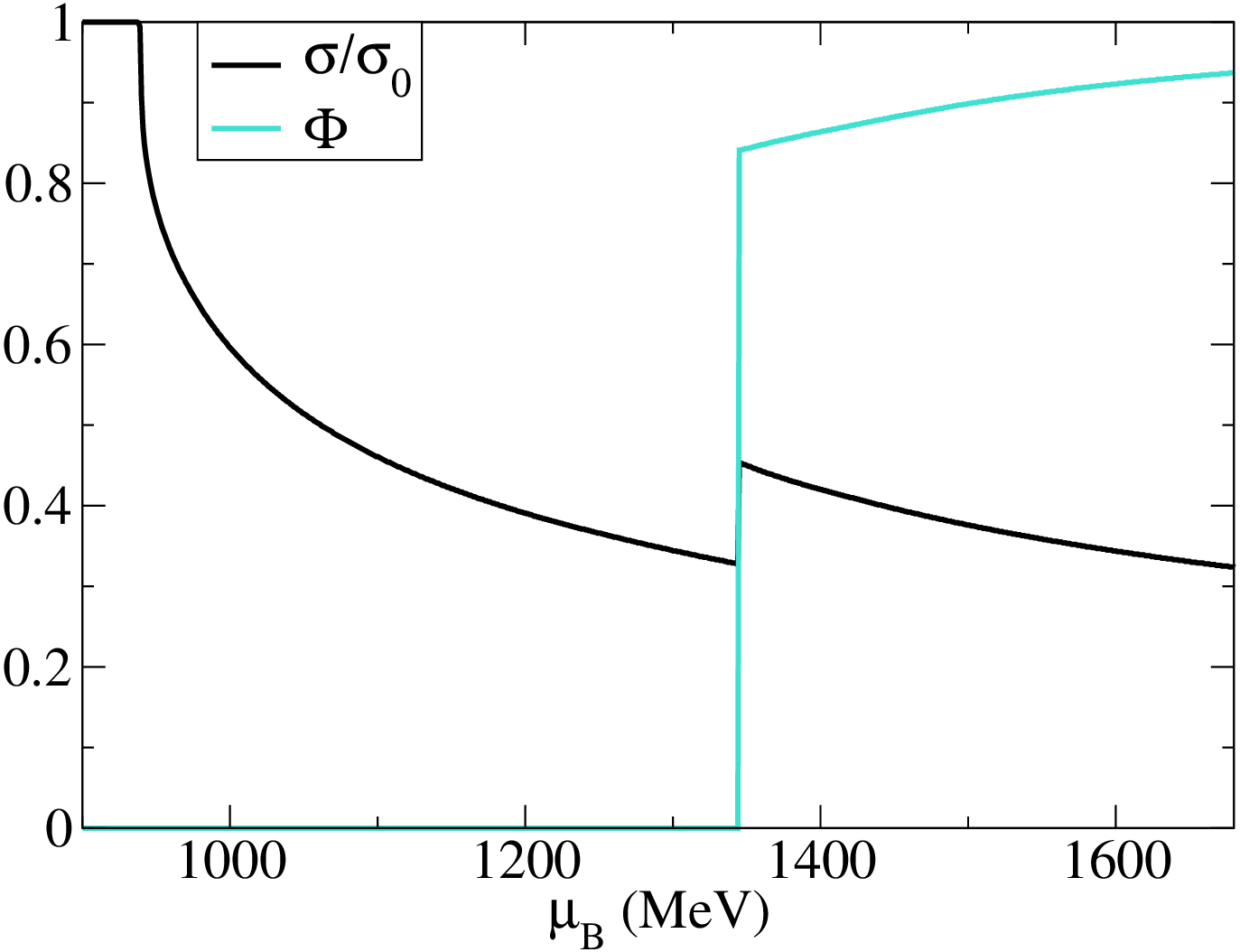}
\caption{\label{Pol}(Color online) Order parameters for chiral symmetry restoration and deconfinement
to quark matter as a function of baryonic chemical potential for star matter at zero temperature.}
\end{figure}

The mesons are treated as classical fields within the mean-field approximation \cite{MFT}. Finite-temperature calculations include the heat bath of hadronic and quark quasiparticles within the grand canonical potential of the system.  It is the defined as:
\begin{eqnarray}
&\frac{\Omega}{V}=-L_{Int}-L_{Self}-L_{SB}-L_{Vac}\nonumber &\\ &\mp T\sum_i \frac{\gamma_i}{(2 \pi)^3}
\int_{0}^{k_{F_i}} \,
d^3k \, \ln(1\pm e^{-\frac{1}{T}(E_i^*(k)-\mu_i^*)}),&
\end{eqnarray}
where $L_{Vac}$ in the vacuum energy, $\gamma_i$ the fermionic degeneracy, $E_{i}^* (k) = \sqrt{k^2+{M^*_i}^2}$ the single particle effective energy, and $\mu_i^*=\mu_i-g_{i\omega} \omega-g_{\phi} \phi- g_{i\rho}\tau_3 \rho$
the effective chemical potential of each species. The chemical potential for each species $\mu_i$ comes from the chemical equilibrium conditions. Finite temperature calculations also include a gas of free pions and kaons. As they have very low mass, they dominate the low density/ high temperature regime. All calculations were performed considering zero net strangeness except the zero temperature star matter case since, for neutron stars,  the time scale is large enough for strangeness not to be conserved.

The effective masses of the baryons and quarks
are generated by the scalar mesons except for a small explicit
mass term $M_0$ (equal to $150$ MeV for nucleons, $354$ MeV for hyperons, $5$ MeV for up and down quarks and $150$ MeV for strange quarks) and the term containing $\Phi$:
\begin{eqnarray}
&M_{B}^*=g_{B\sigma}\sigma+g_{B\delta}\tau_3\delta+g_{B\zeta}\zeta+M_{0_B}+g_{B\Phi} \Phi^2,&
\label{6}
\end{eqnarray}
\begin{eqnarray}
&M_{q}^*=g_{q\sigma}\sigma+g_{q\delta}\tau_3\delta+g_{q\zeta}\zeta+M_{0_q}+g_{q\Phi}(1-\Phi).\nonumber&\\&
\label{7}
\end{eqnarray}
With the increase of temperature/density, the $\sigma$ field
(non-strange chiral condensate) decreases its value, causing
the effective masses of the particles to decrease towards
chiral symmetry restoration. The field $\Phi$ assumes non-zero
values with the increase of temperature/density and, due to its
presence in the baryons effective mass (Eq.~(\ref{6})),
suppresses their presence. On the other hand, the presence of
the $\Phi$ field in the effective mass of the quarks, included
with a negative sign (Eq.~(\ref{7})), insures that they will
not be present at low temperatures/densities. As can be seen
from the different orders of $\Phi$ and different signs in the
new effective mass terms, the motivation for this construction
is not derived from QCD. It is  a simple effective way to
change degrees of freedom within the same model. Note that in
the PNJL approach the coupling of the quarks to the Polyakov
loop can be derived to be included in the quark and antiquark
distribution functions in the grand canonical potential.
However, this leads to non-vanishing quasi-quark contributions
at any temperature below $T_c$, which we avoid in our
phenomenological approach (Eqs.~(6,7)).

\begin{figure}[t]
\centering
\vspace{1.0cm}
\includegraphics[width=8.5cm]{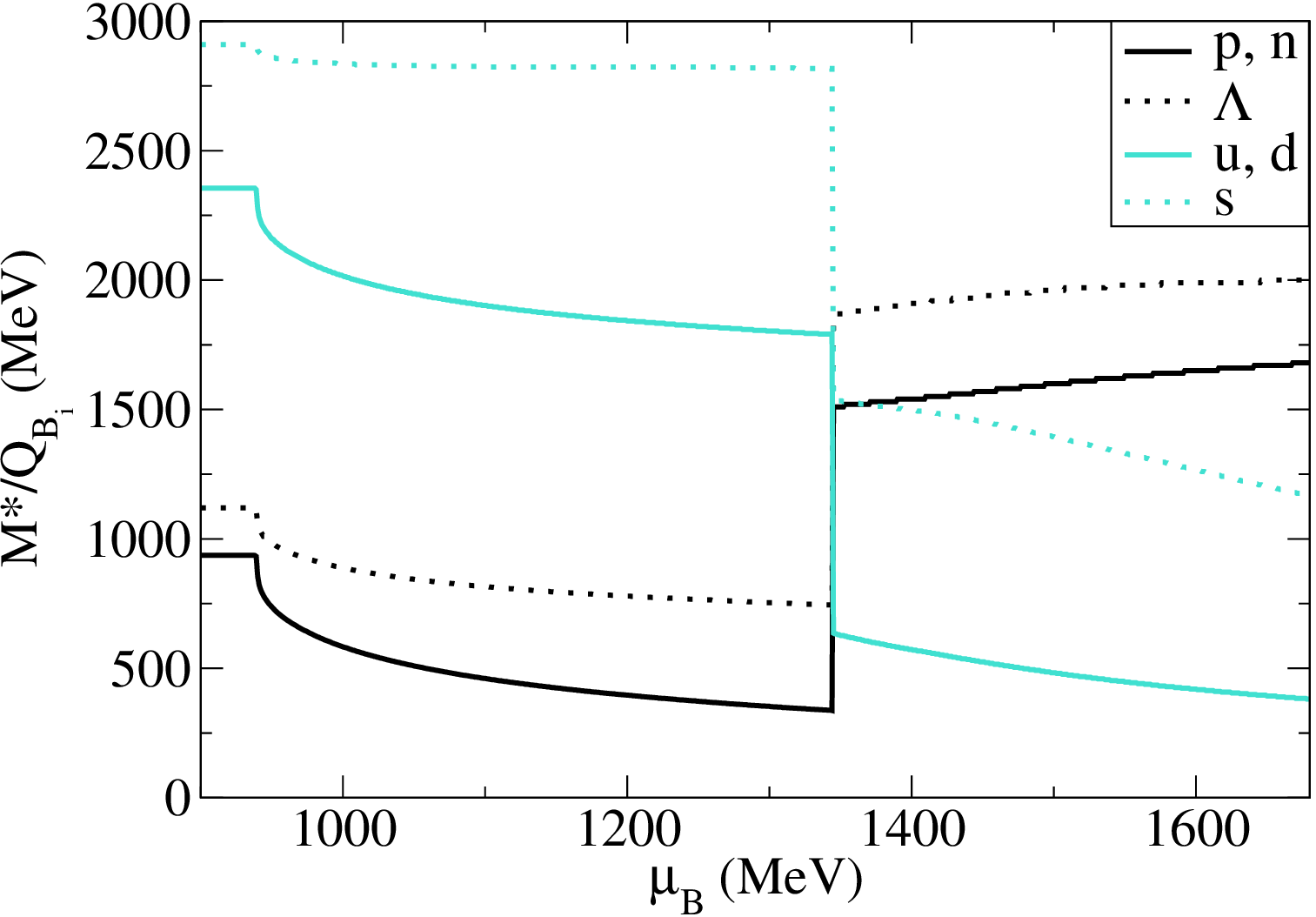}
\caption{\label{Meff}(Color online) Effective normalized mass of different species as a function of baryonic chemical potential for star matter at zero temperature.}
\end{figure}

The behavior of the order parameters of the model is shown in Fig.~\ref{Pol} for neutron
star matter at zero temperature. The difference between this kind of matter and the so-called
symmetric matter comes from the assumption of charge neutrality, essential for the stability of neutron stars, and beta equilibrium.
In this case, the chiral symmetry restoration, which is a crossover for purely hadronic matter,
turns into a first order phase transition by the influence of the strong first order transition
to deconfined matter. The model is consistent in the sense that both order parameters are related.
The small increase in the chiral condensate value during the transition is due to the
smaller quark baryon number (1/3) compared to the baryonic one.

The effective normalized masses of baryons and quarks show the relation
between this quantities and the order parameters, responsible for the
dynamics of the model (Fig.~\ref{Pol} and \ref{Meff}).  Since the coupling constants in the $\Phi$ term of the effective mass formulas are high but still finite, the effective masses of the degrees of freedom not effectively present in each phases are high but also finite. The effective masses normalized by the baryonic number are shown in Fig.~\ref{Meff}.  These quantities are directly related to the onset of particles appearance in the system, that reads:
\begin{eqnarray}\label{limit}
&\mu_{B_{onset}}=\frac{M_{i}^*}{Q_{B_i}}+\mu_e\frac{Q_i}{Q_{B_i}}-\mu_S\frac{Q_{S_i}}{Q_{B_i}}+\frac{g_{i\omega}\omega+g_{i\phi}\phi+g_{i\rho}\tau_3\rho}{Q_{B_i}},\nonumber&\\&
\end{eqnarray}
where $Q_{B_i}$ is the baryonic number, $\mu_e$ is the electron chemical potential, $Q_i$ is the electric charge, $\mu_S$  the strange chemical potential, and $Q_{S_i}$ the strangeness of each species.

\begin{table}
\caption{\label{tabela2}
Additional coupling constants for the model containing baryons and quarks}
\begin{ruledtabular}
\begin{tabular}{ccc}
$ g_{q\omega}=0 $&$ g_{q\phi}=0 $&$ g_{q\rho}=0 $\\
$ g_{q\sigma}=-3.00 $&$ g_{q\delta}=0 $&$ g_{q\zeta}=-3.00 $\\
$ a_0=-1.85 $&$ a_1=-1.44$x$10^{-3} $&$ a_2=-0.08 $\\
$ a_3=-0.40 $&$ g_{B\Phi}=1500$ MeV &$g_{q\Phi}=500$ MeV\\
\end{tabular}
\begin{tabular}{cc}
$\ \  \ T_0=200$ MeV & $ T_0=270$ MeV for pure gauge case\ \ \ \ \  \\
\end{tabular}
\end{ruledtabular}
\end{table}

\begin{figure}[t]
\centering
\vspace{1.0cm}
\includegraphics[width=8.7cm]{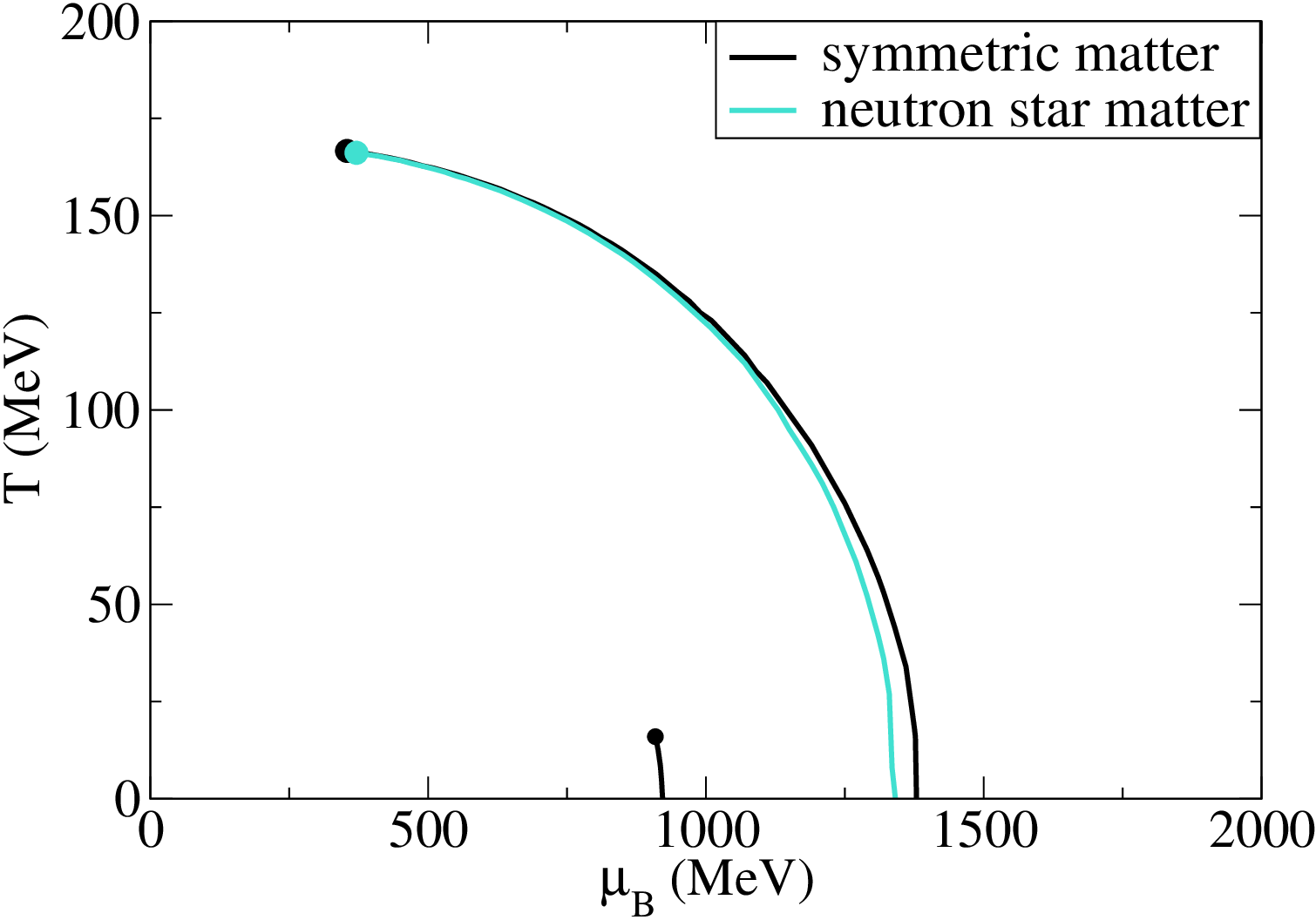}
\caption{\label{phase}(Color online) Phase diagram: temperature versus baryonic chemical potential. The lines represent first order transitions.
The circles mark the critical end-points.}
\end{figure}

Continuing the analogy to the PNJL model, the potential $U$ for $\Phi$ reads:
\begin{eqnarray}
&U=(a_0T^4+a_1\mu^4+a_2T^2\mu^2)\Phi^2&\nonumber\\&+a_3T_0^4\log{(1-6\Phi^2+8\Phi^3-3\Phi^4)}.&
\end{eqnarray}

It is a simplified version of the potential used in
\cite{Ratti1,Ratti2} and adapted to also include terms that
depend on the chemical potential. The two extra terms (that
depend on the chemical potential) are not unique, but the most
simple natural choice whose parameters are chosen to reproduce
the main features of the phase diagram at finite densities.
While in the first part of the potential $\Phi^2$ ensures $U$
to be bound from below, in the second part of the potential
$T_0^4\log{(1-6\Phi^2+8\Phi^3-3\Phi^4)}$ ensures that $\Phi$ is
always (for any region of the phase diagram) bound between zero
and one.

The coupling constants for the quarks are shown in Tab.~\ref{tabela2} and are
chosen to reproduce lattice data as well as known information about the phase diagram. The lattice data includes a first order phase transition at $T=270$ MeV and a pressure functional P(T) similar to Refs. \cite{Ratti1,Ratti2} at $\mu=0$ for pure gauge, a crossover at vanishing chemical potential with a transition temperature of $171$ MeV (determined as the peak of the change of the chiral condensate and $\Phi$) and the location of the critical end-point (at $\mu_c=354$ MeV , $T_c=167$ MeV for symmetric matter in accordance with one of the existent calculations \cite{fodor}). The phase diagram information includes a continuous first order phase transition line that terminates on the zero temperature axis at four times saturation density.

As can be seen in Fig.~\ref{phase}, the transition from hadronic to quark matter obtained is a crossover for small chemical potentials.
 Beyond the critical end-points a first order transition lines for symmetric as well as for star matter begin. The critical temperatures for chiral symmetry
restoration coincide with the ones for deconfinement in both cases. Since the model is able to reproduce nuclear matter
saturation at realistic values for the saturation density, nuclear binding energy, as well as the compressibility and
asymmetry energy, we also show a line in the phase diagram for the nuclear matter liquid-gas phase transition.

\begin{figure}[t]
\centering
\vspace{1.0cm}
\includegraphics[width=8.48cm]{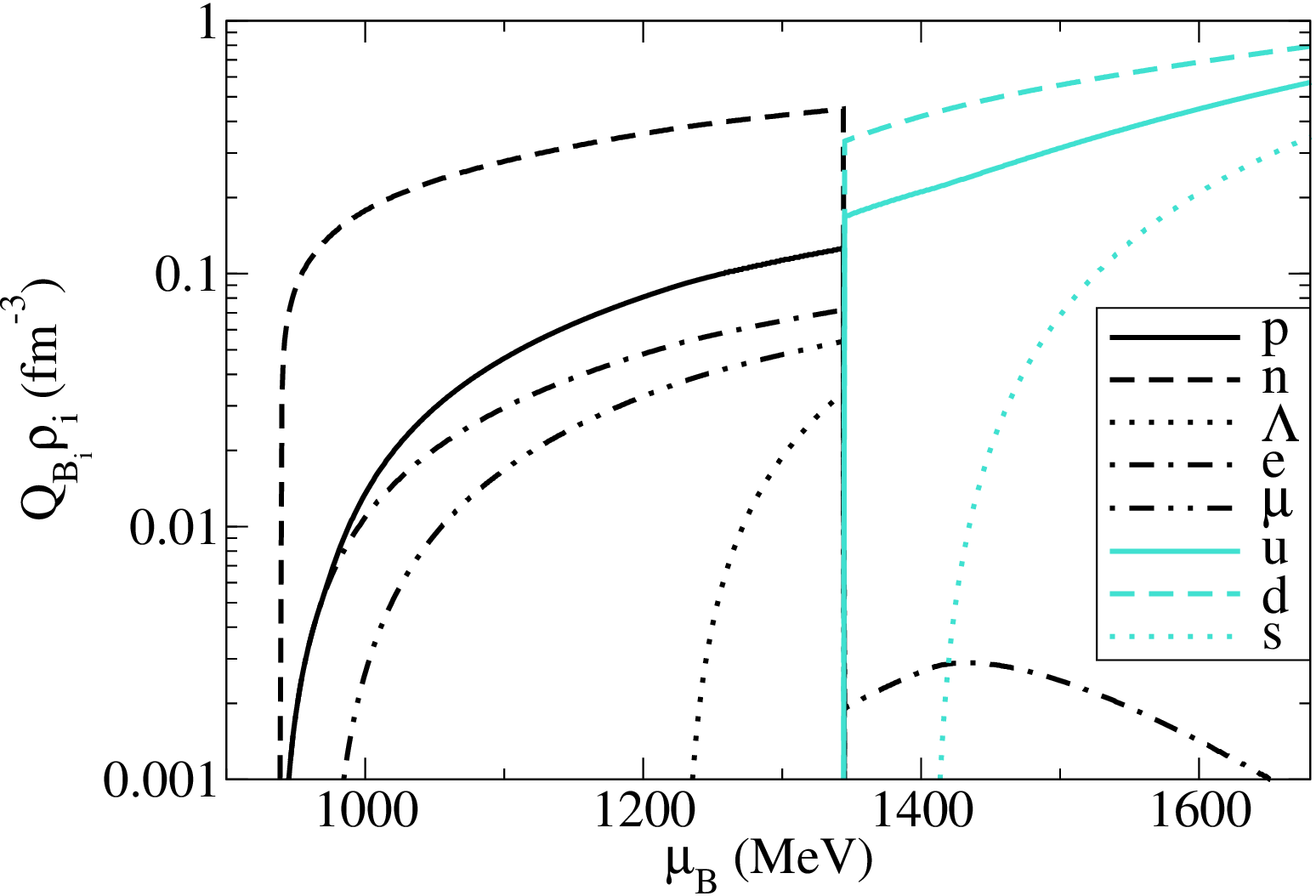}
\caption{\label{popmax}(Color online) Population (baryonic density for different species as a function of baryonic chemical potential) for star matter at zero temperature using local charge neutrality}
\end{figure}

One way to test the model and to compare its predictions with known observational data is to study the high density/low temperature part of
the phase diagram and compare our results with neutron star observations. The critical point for star matter lies at a slightly higher
chemical potential than for the symmetric case and the first order transition line terminates on the zero temperature axis
at $\mu_B=1345$\,MeV (grey line in Fig.~\ref{phase}).
Up to this point, the charge neutrality was considered to be local, meaning that each phase had to be charge neutral by itself.
At finite temperature the two phases contain mixtures of hadrons and quarks, which are dominated by hadrons or quarks,
depending on the respective phase. At vanishing temperature there is no mixture, i.e.
the system exhibits a purely hadronic and purely quark phase (Fig.~\ref{popmax}). The density of
electrons and muons is significant in the hadronic phase but not in the quark phase. The reason for this behavior
is that because the down and strange quarks are also negatively charged,
there is no necessity for the presence of electrons to generate charge neutrality, and only a small amount of leptons
remains to assure beta equilibrium.

\begin{figure}[t]
\centering
\vspace{1.0cm}
\includegraphics[width=8.4cm]{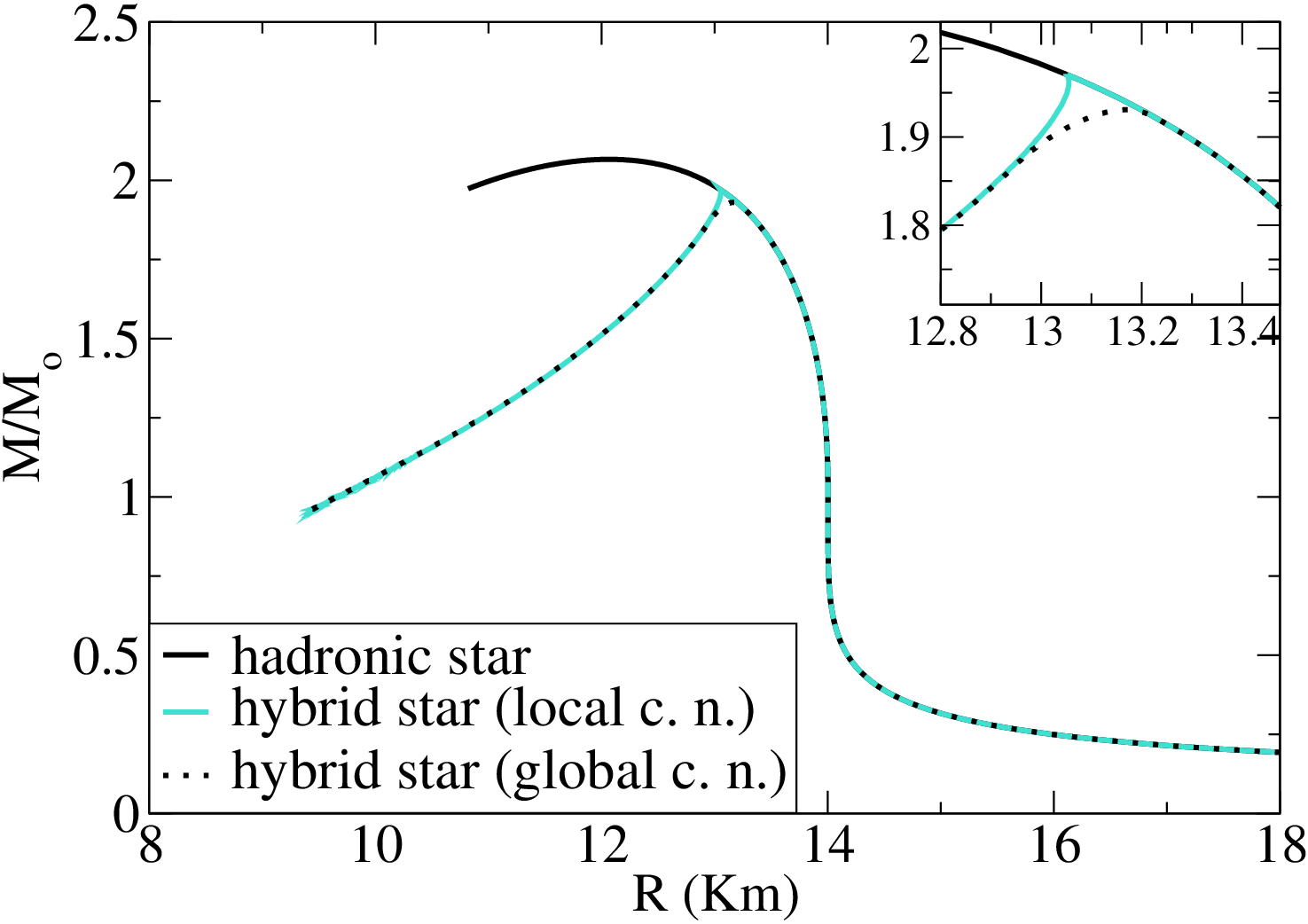}
\caption{\label{mass}(Color online) Mass-radius diagram}
\end{figure}

The quarks are totally suppressed in the hadronic phase and the hadrons are suppressed in the quark phase until a certain chemical potential (above $1700$ MeV for $T=0$). This behavior comes from the fact that  the coupling constants in the $\Phi$ term of the effective mass formulas are high but still finite, so at very high chemical potential the threshold in Eq.~(\ref{limit}) can be reached a second time for hadrons. This threshold, that is higher than the density in the center of neutron stars, establishes a limit for the applicability of the model.
The hyperons, in spite of being included in the calculation, are suppressed by the
appearance of the quark phase. Only a very small amount of $\Lambda$ appears right
before the phase transition (Fig.~\ref{popmax}). The strange quarks appear after the other quarks
and also do not make substantial changes in the system.

The possible neutron star masses and radii are calculated solving the Tolmann-Oppenheimer-Volkof equations \cite{tov1,tov2}.
The solutions for hadronic (same model but without quarks) and hybrid stars are shown in Fig.~\ref{mass}, where
besides our equation of state for the core, a separate equation of state was used for the
crust \cite{crust}. The maximum mass supported against gravity in our model is $2.1M_\odot$ in the
first case and around $2.0M_\odot$ in the second. Because the equation of state for quark matter is much softer than the one for hadronic matter, the star becomes unstable right when the central density is higher than the phase transition threshold.

\begin{figure}[t]
\centering
\vspace{1.0cm}
\includegraphics[width=8.45cm]{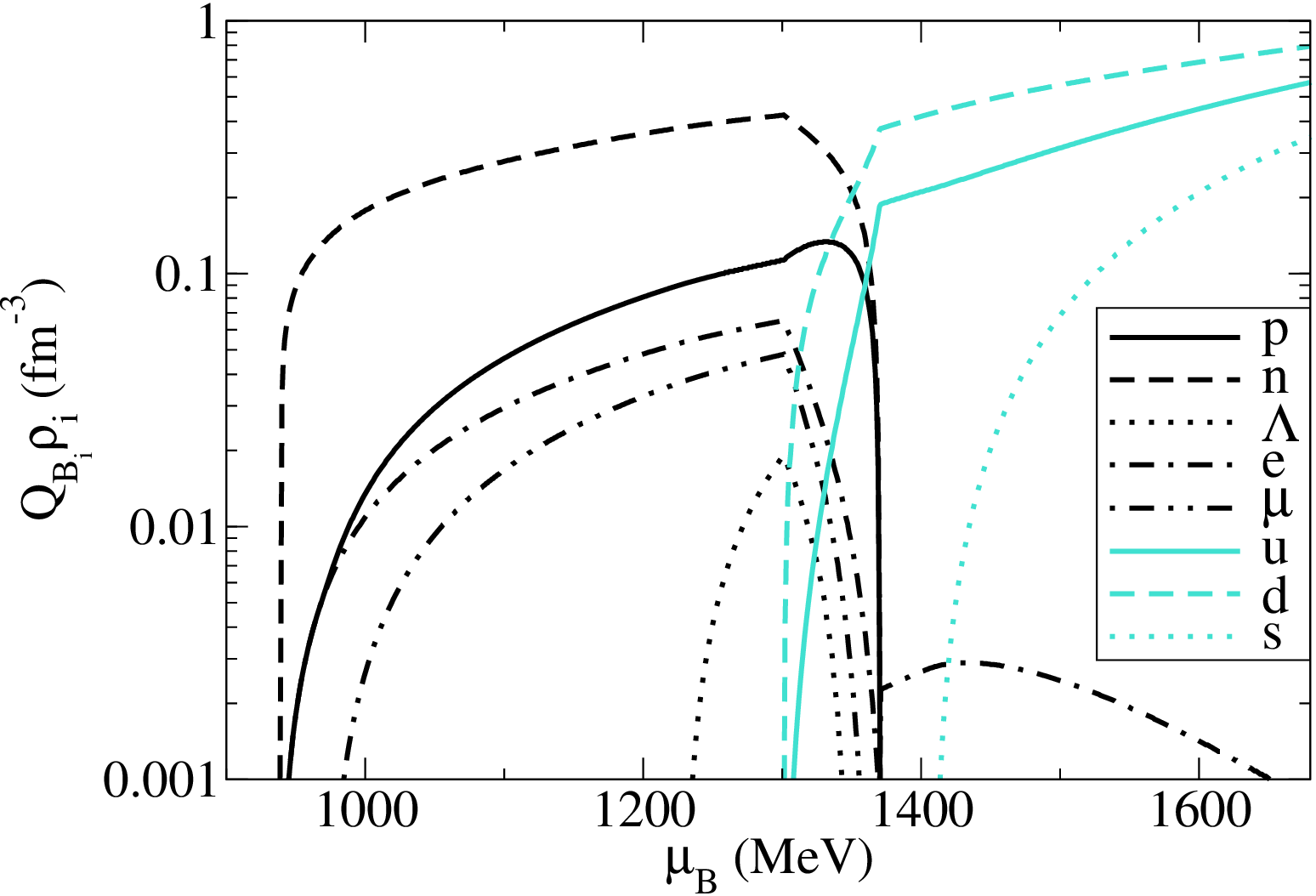}
\caption{\label{popgib}(Color online) Population (baryonic density for different species as a function of baryonic chemical potential) for star matter at zero temperature using global charge neutrality.}
\end{figure}

\begin{figure}[t]
\centering
\vspace{1.0cm}
\includegraphics[width=8.45cm]{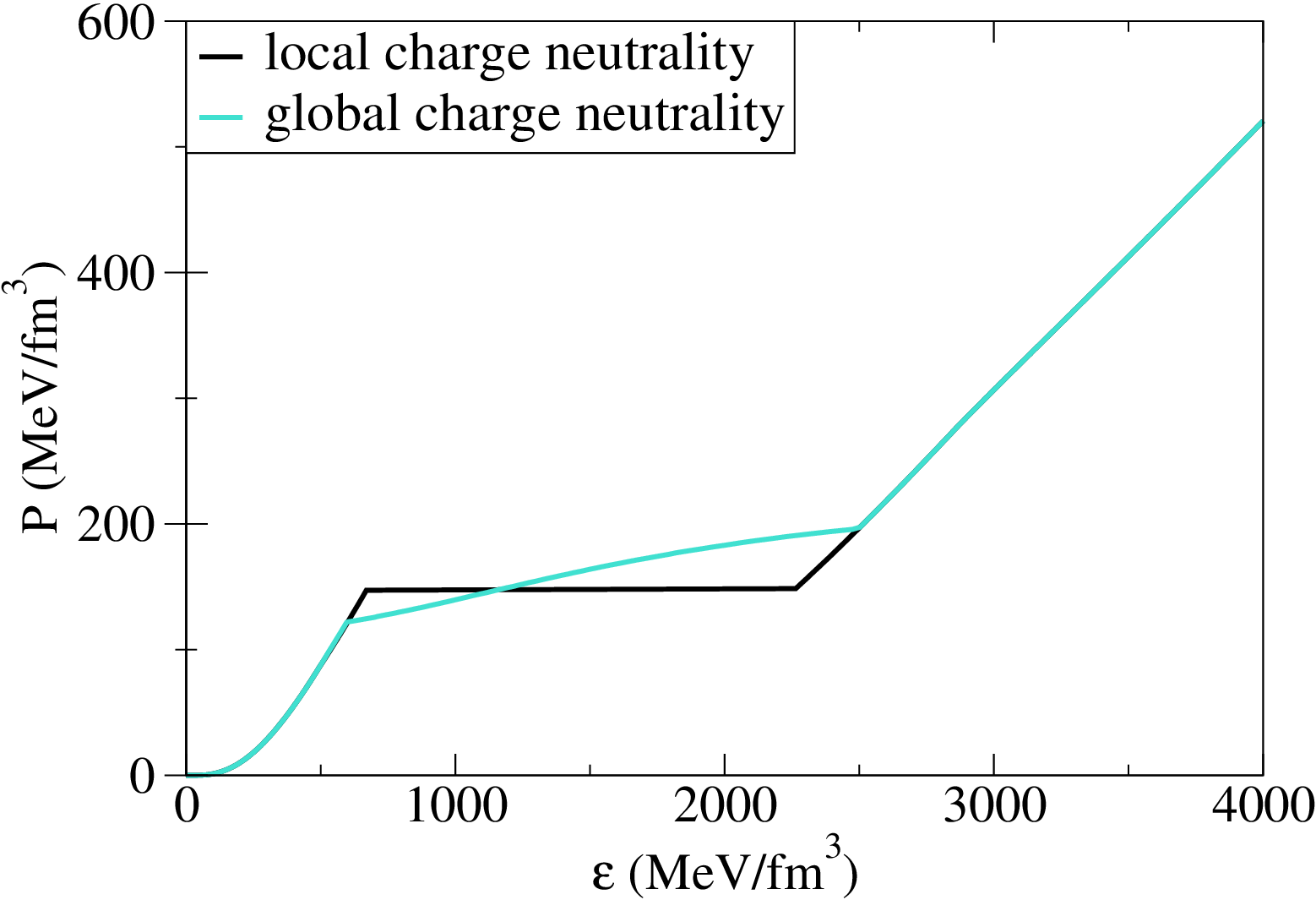}
\caption{\label{eos}(Color online) Equation of State (pressure as a function of energy density) for star matter at zero temperature using local and global charge neutrality.}
\end{figure}

There is still another possible option for the configuration of the particles in the neutron
star \cite{Glendenning:1992vb}. If instead of local we consider global charge neutrality,
we find a mixture of phases. This possibility, which is a more realistic approach, changes the particle
densities in the coexistence region making them appear and vanish in a
smoother way (Fig.~\ref{popgib}). Therefore, the maximum mass allowed for the
star is slightly lower in this case than in the previous one, as can be seen
from the dotted line in Fig.~\ref{mass}; however, this possibility allows stable hybrid stars with a small amount of quarks. The mixed phase constitutes the inner core of the star up to a radius of
approximately 2km. The equation of state for both cases is shown in Fig.~\ref{eos}. The large jump in the pressure in the local charge neutrality case explains why the neutron stars become immediately unstable after the phase transition in this configuration.

We conclude that our model is suitable for the description of neutron stars. The maximum mass predicted
is around the most massive pulsars observed \cite{1,2,maxim,3,4}. The radii lie in the allowed range being practically the same for
hadronic or hybrid stars. A major advantage of our work compared to other studies of hybrid stars
is that because we have only one equation of state for different degrees of freedom we
can study in detail the way in which chiral symmetry is restored and the way deconfinement
occurs at high temperature/density. Since the properties of the physical system, e.g. the
density of particles in each phase, are directly connected to the order parameter for deconfinement $\Phi$ it is not surprising
that we obtain different results in a combined description of the degrees of freedom compared to a simple
matching of two separate equations of state.

Since the model additionally shows a realistic structure of the phase transition
over the whole range of chemical potentials and temperatures as well as phenomenologically acceptable
results for saturated nuclear matter, this approach presents an ideal tool for the
study of ultrarelativistic heavy-ion collisions. Calculations along this line are in progress \cite{soon}.


\begin{thebibliography}{99}

\bibitem{Glendenning:1991ic}
  N.~K.~Glendenning, F.~Weber and S.~A.~Moszkowski,
  Phys.\ Rev.\  C {\bf 45}, 844 (1992).

\bibitem{Weber:1989uq}
  F.~Weber and M.~K.~Weigel,
  Nucl.\ Phys.\  A {\bf 505}, 779 (1989).

\bibitem{Schaffner:1995th}
  J.~Schaffner and I.~N.~Mishustin,
  Phys.\ Rev.\  C {\bf 53}, 1416 (1996)
  [arXiv:nucl-th/9506011].

\bibitem{chiral2}
  P.~Papazoglou, D.~Zschiesche, S.~Schramm, J.~Schaffner-Bielich, H.~Stocker and W.~Greiner,
  Phys.\ Rev.\  C {\bf 59}, 411 (1999).

\bibitem{Heide:1993yz}
  E.~K.~Heide, S.~Rudaz and P.~J.~Ellis,
  Nucl.\ Phys.\  A {\bf 571}, 713 (1994)
  [arXiv:nucl-th/9308002].

\bibitem{Carter:1995zi}
  G.~W.~Carter, P.~J.~Ellis and S.~Rudaz,
  Nucl.\ Phys.\  A {\bf 603}, 367 (1996)
  [Erratum-ibid.\  A {\bf 608}, 514 (1996)]
  [arXiv:nucl-th/9512033].

\bibitem{Bonanno:2008tt}
  L.~Bonanno and A.~Drago,
  arXiv:0805.4188 [nucl-th].

\bibitem{bag0} F. Weber, Prog. Part. and Nucl. Phys. {\bf 54}, 193 (2005), and references therein.
%

\bibitem{Bub}
  M.~Buballa,
  Phys.\ Rept.\  {\bf 407}, 205 (2005)
  [arXiv:hep-ph/0402234].


\bibitem{hybrid1}
  H.~Heiselberg, C.~J.~Pethick and E.~F.~Staubo,
  Phys.\ Rev.\ Lett.\  {\bf 70}, 1355 (1993).

\bibitem{PNJL}
  K.~Fukushima,
  Phys.\ Lett.\  B {\bf 591}, 277 (2004)
  [arXiv:hep-ph/0310121].

\bibitem{chiral1}
  P.~Papazoglou, S.~Schramm, J.~Schaffner-Bielich, H.~Stocker and W.~Greiner,
  Phys.\ Rev.\  C {\bf 57}, 2576 (1998).

\bibitem{eu}
  V.~Dexheimer and S.~Schramm,
  Astrophys.\ J.\ {\bf 683}, 943 (2008).

  \bibitem{MFT} J.D. Walecka, {\sl Theoretical Nuclear And Subnuclear Physics} World Scientific Publishing Company; 2nd edition (2004).

\bibitem{Ratti1}
  C.~Ratti, M.~A.~Thaler and W.~Weise,
  Phys.\ Rev.\  D {\bf 73}, 014019 (2006)

\bibitem{Ratti2}
  S.~Rossner, C.~Ratti and W.~Weise,
  Phys.\ Rev.\  D {\bf 75}, 034007 (2007)

\bibitem{fodor}
  Z.~Fodor and S.~D.~Katz,
  JHEP {\bf 0404}, 050 (2004)
  [arXiv:hep-lat/0402006].

\bibitem{tov1}
  R.~C.~Tolman,
  Phys.\ Rev.\  {\bf 55}, 364 (1939).

\bibitem{tov2}
  J.~R.~Oppenheimer and G.~M.~Volkoff,
  Phys.\ Rev.\  {\bf 55}, 374 (1939).

\bibitem{crust}
 G.~Baym, C.~Pethick and P.~Sutherland,
  Astrophys.\ J.\  {\bf 170}, 299 (1971).

\bibitem{Glendenning:1992vb}
  N.~K.~Glendenning,
  Phys.\ Rev.\  D {\bf 46}, 1274 (1992).

  
    \bibitem{1}
    D.~Barret, J.~F.~Olive and M.~C.~Miller,
  Mon.\ Not.\ Roy.\ Astron.\ Soc.\  {\bf 361}, 855 (2005)
  [arXiv:astro-ph/0505402].
  
  \bibitem{2}
    F.~Ozel,
  Nature {\bf 441}, 1115 (2006).
  
\bibitem{maxim}
  D.~J.~Champion {\it et al.},
  arXiv:0805.2396 [astro-ph].
  
  \bibitem{3}
    J.~Casares, J.~I.~G.~Hernandez, G.~Israelian and R.~Rebolo,
  arXiv:0910.4496 [astro-ph.GA].
  
    \bibitem{4}
      T.~Guver, F.~Ozel, A.~Cabrera-Lavers and P.~Wroblewski,
  arXiv:0811.3979 [astro-ph].
  
  

\bibitem{soon}
J.~Steinheimer, V.~Dexheimer, H.~Petersen, M.~Bleicher, S.~Schramm and H.~Stoecker,
  arXiv:0905.3099 [hep-ph].

\end{thebibliography}
\end{document}